\documentstyle[axodraw]{article}

\renewcommand{\slash}[1]{\rlap/#1}

\renewcommand{\vec}[1]{\mbox{\boldmath $#1$}}

\title{
\centerline{\normalsize hep-ph/9812268 \hfill SINP/TNP/98-34}
\bf Photogravitational processes}

\author{Jos\'e F. Nieves\\
Department of Physics, University of Puerto Rico \\
	R\'{\i}o Piedras, Puerto Rico 00939\\[12pt]
Palash B. Pal\\ 
Saha Institute of Nuclear Physics, 1/AF Bidhan-Nagar \\ 
	Calcutta 700064, INDIA}

\date{December 1998}

\begin{document}
\maketitle

\begin{abstract} 

In the linearized theory of gravity coupled to the electromagnetic field, 
we calculate the cross sections
of two processes in which both the gravitational and electromagnetic
interactions play a role. Those processes, 
in which a graviton is produced in the final state,
are:
(a) the photoproduction of graviton off electrons, and (b) the production
of a photon and a graviton from $e^+$-$e^-$ pair annihilation.
The motivation and outlook for these calculations are discussed.

\end{abstract}

\section{Introduction}
\setcounter{equation}{0}
\label{s:intro}
In the framework of Quantum Field Theory, 
the various fundamental interactions are
understood as due to the exchange of gauge bosons.
In the case of gravity,
at least for weak gravitational fields, the
linearized version of Einstein's theory of general relativity can be
used to construct a quantum theory involving the spin-2 graviton which
mediates the gravitational interactions.

Historically however, the interactions
have been discovered by observing processes involving
fermions, and not the gauge bosons, 
in the outer states. The gauge particles 
act as virtual mediators in such processes.  At some later stage, the
processes where the mediating particles also appear in
the initial or the final state have been discovered. 
That was the case of
electromagnetism, for example, 
where the interaction between electrons had been known for a
long time, but Compton scattering and the photoelectric effect were
discovered in the twentieth century. Similarly,
while the weak interaction between
fermions was known since the beginning of this century, the processes
involving real $W$ and $Z$ bosons have been observed only very
recently.

It seems to be a natural extension to this,
to inquire about processes involving gravitons in the initial
and/or the final state.
As examples of such processes, 
in this article we consider two of them,
which involve one graviton and a photon, and to which refer as 
``photogravitational processes''.
The first one is the photoproduction of graviton
off electrons and the other one is the production of a photon and a
graviton from $e^+e^-$  annihilation.

Besides the intrinsic interest that the study of these processes
have, there are some practical motivations for 
this undertaking that we should mention.

A few years ago we showed that, in a medium that 
contains electrons but not the other charged leptons,
such as normal matter, the electromagnetic interactions of neutrinos are
not the same for all the neutrino flavors\cite{dnp:nuem}. 
It was
observed there that, in the presence of a static
magnetic field, the effective electromagnetic interactions of the
neutrinos produce an additional contribution to the neutrino index of
refraction which modifies the condition for resonant oscillations in
matter\cite{others:nuem}.  This effect is the basis for
the explanation
of the large birth velocities of pulsars in terms of the 
asymmetric emission of neutrinos from the cooling protostar,
which is produced by the matter-enhanced
neutrino oscillations biased by the supernova's magnetic 
field~\cite{ks}.

Recently\cite{np:gravnu} we have shown that,
in analogy with the fact that in
a matter background 
the electron neutrinos have different electromagnetic
interactions than the muon or tau neutrinos,
their gravitational interactions also differ.
Furthermore, motivated by the fact that the gravitational
interactions of neutrinos may be important in some 
contexts in neutrino physics\cite{hl,piriz}, 
in Ref.\ \cite{np:gravnu}  we determined
the effective gravitational
interactions of neutrinos in a matter background,
by calculating the one-loop contribution to the
neutrino stress-energy tensor, which is the gravitational analog
of the electromagnetic current.
Those calculations  were based
on the linearized theory of gravitational couplings
of fermions, including the interaction terms with
the $W$ and $Z$ gauge bosons, and an outline of their derivation was
presented in that reference. In addition, special attention was given
to show that the result for the matter-induced gravitational vertex 
of the neutrino is (gravitational) gauge invariant.

In an outgrowth and extension of the work
of Ref.\ \cite{np:gravnu} which is currently underway,
the electromagnetic interactions, in addition to the
gravitational ones, play a role.  In that case,
the result of the calculation should
be gravitational and electromagnetic gauge invariant.
In order to understand some
subtle issues that have to do with the graviton couplings
to the photon and (charged) fermions,
we found convenient and instructive to
consider processes involving both
the gravitational and the electromagnetic interactions,
but isolated from the additional complications that arise in the
calculations of effective interactions in a matter background.
The photogravitational processes that we have considered here,
besides their intrinsic interest, serve the stated purpose well.

We should mention also that the processes that we consider here
are not related to the soft graviton processes that are
considered in connection with the equivalence principle and
the soft graviton theorem\cite{softgravitons}.  Those are
processes in which the graviton is radiated by one of the external
particles, but the process  can occur also without the radiated
graviton.  In contrast, the processes that we consider
cannot occur in the absence of graviton emission.

We have organized the presentation as follows.  In Section\ \ref{s:int}
we outline the derivation of the various gravitational
couplings that are required in the calculation. 
The calculation of the cross section for the
photoproduction of gravitons is carried out in Section\ \ref{s:photprod},
where special attention is given to check that the couplings
derived in Section\ \ref{s:int} ensure a gauge invariant result.
Similarly, in Section\ \ref{s:pairann} we consider the process of
$e^+e^-$ pair annihilation, and finally Section\ \ref{s:conc}
contains our concluding remarks.

\section{Interactions involving gravitons}
\setcounter{equation}{0}
\label{s:int}
In this section, we derive the various couplings involving the graviton
which will be relevant for us. As already indicated, all the results
are derived in linearized theory of gravity, in which the metric
tensor is written as
\begin{eqnarray}
g_{\lambda\rho} = \eta_{\lambda\rho} + 2\kappa h_{\lambda\rho}
\label{defh}
\end{eqnarray}
where $\eta_{\lambda\rho}$ is the flat space metric. We then expand the
Lagrangian in the presence of gravity in powers of $\kappa$ and keep only
the first order terms. In this formulation, $h_{\lambda\rho}$ appears as
the graviton field, which is a spin-2 quantum field coupled to the
stress-energy tensor, whose interactions can be studied in the flat
Minkowskian background. The Einstein-Hilbert action for pure gravity is
given by
\begin{eqnarray}
{\cal A} = {1\over 16\pi G} \int d^4x\; \sqrt{-{\tt g}} \; R \,,
\end{eqnarray}
where $R$ is the Ricci scalar, $\tt g$ is the determinant of the
matrix $g_{\lambda\rho}$, and $G$ is the Newton's constant. Using Eq.\
(\ref{defh}), we can verify that this gives the correct kinetic terms
for the spin-2 field if we make the identification
\begin{eqnarray}
\kappa = \sqrt{8\pi G} \,.
\end{eqnarray}
We now discuss the couplings of the graviton to fermions
and the photon.

\subsection{Fermion couplings}

These couplings were already considered in 
Ref.~\cite{np:gravnu}, and we
refer to that work for the details of the derivation. 
The final result is that the
coupling of the graviton field $h_{\lambda\rho}$ with the electron
field can be written as
\begin{eqnarray}\label{Leeh}
{\cal L}^{(ee)}_{h} = -\kappa h^{\lambda\rho} (x) \widehat
T^{(e)}_{\lambda\rho} (x) \,, 
\end{eqnarray}
where the stress-energy tensor operator $\widehat
T_{\lambda\rho}^{(e)}$ for the electron field is given by
\begin{eqnarray}\label{stresstensor}
\widehat T^{(e)}_{\lambda\rho} (x) = \left\{
{i\over 4} \overline \psi(x) \left[\gamma_\mu \partial_\nu + \gamma_\nu
\partial_\mu \right] \psi(x) + {\rm h.c.} \right\} -
\eta_{\lambda\rho} {\cal L}_0^{(e)} (x) \,.
\end{eqnarray}
Here ${\cal L}_0^{(e)}(x)$ is the Lagrangian for the
free Dirac field, which we write in the explicitly Hermitian form
\begin{eqnarray}\label{Lf0}
{\cal L}^{(e)}_0 = \left[
\frac{i}{2}\overline\psi \gamma^\mu \partial_\mu \psi 
+ {\rm h.c.} \right]
- m \overline\psi\psi  \,,
\end{eqnarray}
$m$ being the mass of the electron. 
The Feynman rules derived from this interaction are presented in
Fig.~\ref{f:ffh}, where
\begin{eqnarray}\label{V}
V_{\lambda\rho} (p,p') = \frac14 \left[
\gamma_\lambda(p + p')_\rho + 
\gamma_\rho(p + p')_\lambda \right]
- \frac12 \eta_{\lambda\rho}
\left[(\rlap/ p - m) + (\rlap/ p' - m) \right] \,.
\end{eqnarray}
For fermions on-shell these agree with the results quoted in
standard textbooks\cite{scadron}. However, the additional
terms that appear in Eq.\ (\ref{V}) for 
off-shell fermions are important for us, since there are virtual
fermions in the diagrams that we have to consider.
\begin{figure} 
\begin{center}
\begin{picture}(100,60)(-30,0)
\Photon(0,30)(0,60){2}{5}
\Photon(0,30)(0,60){-2}{5}
\Text(5,50)[l]{$h_{\lambda\rho}$}
\ArrowLine(-30,30)(0,30)
\Text(-15,20)[b]{$p$}
\ArrowLine(0,30)(30,30)
\Text(15,20)[b]{$p'$}
\Text(40,30)[l]{$=-i\kappa V_{\lambda\rho}(p,p')$}
\end{picture}
\end{center}

\caption[]{Feynman rule for the tree-level gravitational
vertex of a fermion.
\label{f:ffh}
}
\end{figure} 

\subsection{Couplings involving photons}
The graviton couplings that involve the photon field are
determined by the Lagrangian density
\begin{eqnarray}
\label{Lem}
{\cal L}_{\rm em} = -\frac{1}{4}\sqrt{-\tt g} \;
F_{\mu\nu}F_{\alpha\beta} g^{\mu\alpha} g^{\nu\beta} +
e \sqrt{- \tt g}\; \overline \psi\gamma^a v_a{}^\mu \psi A_\mu \,,
\end{eqnarray}
where in our notation the electric charge of the electron is $-e$, 
and $v_a{}^\mu$ are the vierbeins (or tetrads) which satisfy the
conditions 
\begin{eqnarray}\label{vierbeins}
\eta^{ab} v_a{}^\mu v_b^\nu =  g^{\mu\nu} \,, \qquad 
g_{\mu\nu} v_a{}^\mu v_b{}^\nu =  \eta_{ab} \,.
\end{eqnarray}
These are necessary also for deriving the interaction in 
Eq.\ (\ref{Leeh}) by expanding in terms of $\kappa$, a procedure which has
been detailed in Ref.~\cite{np:gravnu}. Mimicking that procedure here we
put
\begin{eqnarray}
v_a{}^\mu & \simeq & \eta_a{}^\mu - \kappa h_a{}^\mu \,,
\end{eqnarray}
which follows from Eqs.\ (\ref{defh}) and (\ref{vierbeins}), and
also
\begin{eqnarray}
\sqrt{-{\tt g}} = 1 + \kappa \eta_{\mu\nu} h^{\mu\nu} 
\end{eqnarray}
to first order in $\kappa$.

Substituting these in Eq.\ (\ref{Lem}), we obtain the interaction
terms involving the graviton,
\begin{eqnarray}
\label{Lemh}
{\cal L}^{(\rm em)}_h = {\cal L}^{(eA)}_h + {\cal L}^{(AA)}_h\,,
\end{eqnarray}
where
\begin{eqnarray}
{\cal L}^{(eA)}_h = e\kappa h^{\lambda\rho} a_{\mu\nu\lambda\rho}
\overline\psi\gamma^\nu \psi A^\mu \,,
\label{LeA}
\end{eqnarray}
with
\begin{eqnarray}
a_{\mu\nu\lambda\rho} = \eta_{\mu\nu} \eta_{\lambda\rho} -
\frac12 \left( \eta_{\mu\lambda} \eta_{\nu\rho} + \eta_{\nu\lambda}
\eta_{\mu\rho} \right) \,.
\label{a}
\end{eqnarray}

Notice that, in order to obtain the correct field equations from Eq.\
(\ref{Lem}), it is important that $g^{\mu\nu}$ (or equivalently
$v_a{}^\mu$) and $A_\mu$ are treated as the independent variables and
not, for example, the contravariant vector $A^\mu$. 
If we were to use $A^\mu$ as
the independent variable, the couplings obtained would
not be the correct ones and the resulting amplitudes would not
be gauge invariant.\footnote{Such a
wrong coupling was in fact presented in Ref.~\cite{np:gravnu}. However,
it turns out that the final results derived in that paper are still
correct, although the amplitudes corresponding to the individual diagrams are
not. This will be discussed in detail elsewhere.} 
However, once we
separate out the first order terms in $\kappa$, we can raise and lower
the indices in those terms by the flat metric $\eta_{\mu\nu}$ since we
are interested in the first order terms in the final result. Thus, in
those terms, we can blatantly use upper or lower indices, as we
have done in Eq.\ (\ref{LeA}).

\begin{figure} 
\begin{center}
\begin{picture}(180,60)(0,0)
\Photon(50,30)(100,0){-2}{7}
\ArrowLine(96,8)(86,14)
\Text(105,0)[l]{$A_\mu(k)$}
\Photon(50,30)(100,60){2}{7}
\ArrowLine(86,46)(96,52)
\Text(105,60)[l]{$A_\nu(k')$}
\Photon(0,30)(50,30){2}{7}
\Photon(0,30)(50,30){-2}{7}
\Text(25,35)[b]{$h_{\lambda\rho}$}
\Text(140,30)[l]{$= -i\kappa C_{\mu\nu\lambda\rho}(k,k')$}
\end{picture}
\end{center}

\caption[]{Feynman rule for the coupling of the photon with
the graviton.
\label{f:AAh}
}
\end{figure}
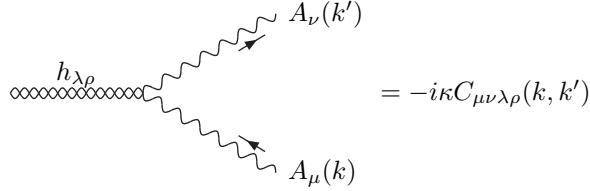 
The photon couplings with the graviton are determined as
\begin{eqnarray}\label{LAAh}
{\cal L}^{(AA)}_{h} = -\kappa h^{\lambda\rho} (x) \widehat
T^{(A)}_{\lambda\rho} (x) \,, 
\end{eqnarray}
where
\begin{eqnarray}
\label{TA}
\widehat T_{\lambda\rho}^{(A)}
& = & \frac14 \eta_{\lambda\rho} F_{\mu\nu} F^{\mu\nu} +
F_{\lambda\mu}F^\mu{}_\rho 
\end{eqnarray}
is the stress-energy tensor operator for the photon field.
The Feynman rule for the photon-photon-graviton vertex is parametrized
in Fig.~\ref{f:AAh}. With the momentum convention shown in
Fig.~\ref{f:AAh}, $C_{\mu\nu\lambda\rho}$ is given by
	\begin{eqnarray}
C_{\mu\nu\lambda\rho} (k,k') &=& \eta_{\lambda\rho} ( \eta_{\mu\nu} k
\cdot k' - k'_\mu k_\nu ) - \eta_{\mu\nu} (k_\lambda k'_\rho +
k'_\lambda k_\rho ) \nonumber\\*
&& + k_\nu (\eta_{\lambda\mu} k'_\rho +
\eta_{\rho\mu} k'_\lambda)
+ k'_\mu (\eta_{\lambda\nu} k_\rho +
\eta_{\rho\nu} k_\lambda) \nonumber\\*
&& - k \cdot k' (\eta_{\lambda\mu} \eta_{\rho\nu} +
\eta_{\lambda\nu} \eta_{\rho\mu})
\label{C}
	\end{eqnarray}
This is the same as what is given in Scadron's book, except for an
overall factor of 2 which we believe is a mistake or misprint in the
book.

\section{Photoproduction of gravitons}
\setcounter{equation}{0}
\label{s:photprod}
\subsection{The amplitude and its invariances}
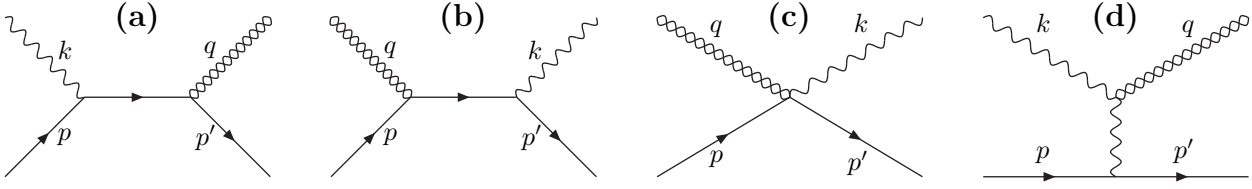
\begin{figure} 
\begin{center}
\begin{picture}(100,60)(0,0)
\ArrowLine(0,0)(30,30)
\Text(20,15)[l]{$p$}
\Text(80,15)[r]{$p'$}
\Text(20,48)[l]{$k$}
\Text(80,48)[r]{$q$}
\ArrowLine(30,30)(70,30)
\ArrowLine(70,30)(100,0)
\Photon(0,60)(30,30){2}{6}
\Photon(70,30)(100,60){2}{6}
\Photon(70,30)(100,60){-2}{6}
\Text(50,60)[]{\large\bf (a)}
\end{picture}
\qquad
\begin{picture}(100,60)(0,0)
\ArrowLine(0,0)(30,30)
\Text(20,15)[l]{$p$}
\Text(80,15)[r]{$p'$}
\Text(20,48)[l]{$q$}
\Text(80,48)[r]{$k$}
\ArrowLine(30,30)(70,30)
\ArrowLine(70,30)(100,0)
\Photon(70,30)(100,60){2}{6}
\Photon(0,60)(30,30){2}{6}
\Photon(0,60)(30,30){-2}{6}
\Text(50,60)[]{\large\bf (b)}
\end{picture}
\qquad
\begin{picture}(100,60)(0,0)
\ArrowLine(0,0)(50,30)
\Text(20,10)[lt]{$p$}
\Text(80,10)[rt]{$p'$}
\Text(20,55)[lb]{$q$}
\Text(80,55)[rb]{$k$}
\ArrowLine(50,30)(100,0)
\Photon(50,30)(100,60){2}{7}
\Photon(0,60)(50,30){2}{7}
\Photon(0,60)(50,30){-2}{7}
\Text(50,60)[]{\large\bf (c)}
\end{picture}
\qquad
\begin{picture}(100,60)(0,0)
\Text(20,7)[lb]{$p$}
\Text(80,7)[rb]{$p'$}
\Text(20,55)[lb]{$k$}
\Text(80,55)[rb]{$q$}
\ArrowLine(0,0)(50,0)
\ArrowLine(50,0)(100,0)
\Photon(50,0)(50,30){2}{4}
\Photon(0,60)(50,30){2}{7}
\Photon(50,30)(100,60){2}{7}
\Photon(50,30)(100,60){-2}{7}
\Text(50,60)[]{\large\bf (d)}
\end{picture}
\end{center}

\caption[]{Tree level diagrams for the photoproduction of
gravitons off electrons. $k$ and $q$ label the lines of the incomming
photon and outgoing graviton, respectively.
\label{f:photprod} 
}
\end{figure}

In this section we consider the process
	\begin{eqnarray}
e^- (p) + \gamma (k) \to e^-(p') + {\cal G}(q) \,,
	\end{eqnarray}
for which the tree level diagrams are shown in Fig.~\ref{f:photprod}.
Writing the Feynman amplitude as
\begin{eqnarray}
\label{defGamma}
iM &  = & (-i\kappa)(ie)
i\left[ \overline u(p')\Gamma_{\mu\lambda\rho} u(p) \right]
\varepsilon^{\ast\lambda\rho} (q) \epsilon^\mu (k) \nonumber\\
& \equiv & ie\kappa{\cal M} \,,
\end{eqnarray}
the contributions to $\Gamma_{\mu\lambda\rho}$ from the various diagrams
are given by
	\begin{eqnarray}
\Gamma^{(a)}_{\mu\lambda\rho} &=& V_{\lambda\rho} (p'+q,p') S_F(p'+q)
\gamma_\mu \\ 
\Gamma^{(b)}_{\mu\lambda\rho} &=& \gamma_\mu S_F(p-q) 
V_{\lambda\rho} (p,p-q) \\ 
\Gamma^{(c)}_{\mu\lambda\rho} &=& \gamma^\alpha
a_{\alpha\mu\lambda\rho} \\  
\Gamma^{(d)}_{\mu\lambda\rho} &=& \gamma_\alpha D^{\alpha\nu}_F(k-q) 
C_{\mu\nu\lambda\rho} (k,k-q)
\nonumber\\*
&=& -\gamma^\nu \Delta(k-q)
C_{\mu\nu\lambda\rho} (k,k-q)  \,,
	\end{eqnarray}
where $a_{\mu\nu\lambda\rho}$ and $C_{\mu\nu\lambda\rho}$ have been
defined in Eqs.\ (\ref{a}) and (\ref{C}), and we have written the
photon propagator as
	\begin{eqnarray} 
D^{\mu\nu}_F (k')
= - \eta^{\mu\nu} \Delta(k') = -\frac{\eta^{\mu\nu}}{k'^2}  \,.
	\end{eqnarray}
As usual, the spinors that appear in Eq.\ (\ref{defGamma})
satisfy
\begin{eqnarray}
\overline u(p') \slash p' &=& m \overline u(p') \label{ubarp/}\\*
\slash p u(p) &=& m u(p) \,.
\label{p/u}
\end{eqnarray}
%

\subsubsection{Checking gravitational gauge invariance}
Gravitational gauge invariance amounts to the statement that
the condition
\begin{eqnarray}
\label{transcond}
q^\lambda \overline u(p')\Gamma_{\mu\lambda\rho} u(p) = 0 
\end{eqnarray}
holds when all particles, other than the graviton, are on-shell. 
In order to verify this, let
us consider first $\Gamma^{(a)}_{\mu\lambda\rho}$. Note that
\begin{eqnarray}
q^\lambda V_{\lambda\rho}(p' + q,p') = 
\frac{1}{4}[\slash{q}(2p' + q)_\rho +
\gamma_\rho(2p' + q)\cdot q] - 
\frac{1}{2}\slash{q}q_\rho \,,
\end{eqnarray}
where we have used Eq.\ (\ref{ubarp/}). We now use
the identities
\begin{eqnarray}
(2p'\cdot q + q^2)\; S_F(p' + q)
= \slash p' + \slash q + m \,,
\end{eqnarray}
and
\begin{eqnarray}
\slash{q} & = & S^{-1}_F(p' + q) - 
S^{-1}_F(p') \nonumber\\
& \rightarrow & S^{-1}_F(p' + q) \,,
\end{eqnarray}
where the last step follows because of Eq.\ (\ref{ubarp/}). This
finally gives
\begin{eqnarray}
\label{qGamma_a}
q^\lambda \Gamma^{(a)}_{\mu\lambda\rho} &=& 
\frac{1}{4}[4p'_\rho - q_\rho +
\gamma_\rho\slash q] \gamma_\mu \,.
\end{eqnarray}
For the diagram (b) the procedure is similar and the result for that
case is
\begin{eqnarray}
\label{qGamma_b}
q^\lambda \Gamma^{(b)}_{\mu\lambda\rho} &=& 
\frac{1}{4} \gamma_\mu [-4p_\rho - q_\rho + \slash q \gamma_\rho] \,. 
\end{eqnarray}
Also, trivially,
\begin{eqnarray}
\label{qGamma_c}
q^\lambda \Gamma^{(c)}_{\mu\lambda\rho} &=& 
\gamma_\mu q_\rho - \frac12 \gamma_\rho q_\mu - \frac12 \eta_{\mu\rho}
\slash q \,.
\end{eqnarray}
Finally, for the diagram (d), we need to use also the on-shell
conditions for the photon
\begin{eqnarray}
\epsilon^\mu(k) k_\mu &=& 0 \,, \nonumber\\*
k^2 &=& 0 \,,
\label{photonshell}
\end{eqnarray}
which imply
\begin{eqnarray}
\gamma_\nu q^\lambda C_{\mu\nu\lambda\rho} (k,k-q) &=& 
\left[ \gamma_\mu k_\rho - \eta_{\mu\rho} \slash k \right] 
\Delta^{-1}(k-q) + 
 (\slash k - \slash q) \left[ q_\mu k_\rho - \eta_{\mu\rho}
k \cdot q \right] \,.
\end{eqnarray}
The last term vanishes between the spinors since $\slash k - \slash
q=\slash p'-\slash p$, so we obtain
\begin{eqnarray}
\label{qGamma_d}
q^\lambda \Gamma^{(d)}_{\mu\lambda\rho} &=& 
-\gamma_\mu k_\rho + \eta_{\mu\rho} \slash k \,.
\end{eqnarray}
Adding Eqs.\ (\ref{qGamma_a}) - (\ref{qGamma_c}) and 
(\ref{qGamma_d}), and using
\begin{eqnarray}
\frac14[
\gamma_\rho\slash q \gamma_\mu + 
\gamma_\mu \slash q \gamma_\rho] = \frac12 [\gamma_\mu q_\rho +
\gamma_\rho q_\mu - \eta_{\mu\rho} \slash q ] \,, 
\end{eqnarray}
we finally obtain
\begin{eqnarray}
q^\lambda \Gamma_{\mu\lambda\rho} &=& \gamma_\mu (p'-p-k+q)_\rho 
+ \eta_{\mu\rho} (\slash k - \slash q) \,, 
\end{eqnarray}
which leads to Eq.\ (\ref{transcond}) because the first term
is zero due to momentum conservation
and the second vanishes when it is multiplied by the spinors.

\subsubsection{Checking electromagnetic gauge invariance}
The amplitude should also be invariant under electromagnetic gauge
transformations. To check this, we note that
\begin{eqnarray}
k^\mu \Gamma^{(a)}_{\mu\lambda\rho} &=& V_{\lambda\rho} (p'+q,p')
S_F(p+k) \slash k \,,
\end{eqnarray}
using the conservation of momentum to write the fermion propagator in terms
of $p+k$ rather than $p'+q$ here.  Writing $\slash k = (\slash
p+\slash k-m)-(\slash p -m)$ and noting that $(\slash p-m)$ vanishes
when it acts on the spinor on the right, we obtain $S_F(p+k) \slash k =1$,
so that 
\begin{eqnarray}
\label{qGamma_a2}
k^\mu \Gamma^{(a)}_{\mu\lambda\rho} &=& V_{\lambda\rho} (p'+q,p') \,.
\end{eqnarray}
Similarly, 
\begin{eqnarray}
\label{qGamma_b2}
k^\mu \Gamma^{(b)}_{\mu\lambda\rho} &=& -V_{\lambda\rho} (p,p-q) \,,
\end{eqnarray}
and, trivially,
\begin{eqnarray}
\label{qGamma_c2}
k^\mu \Gamma^{(c)}_{\mu\lambda\rho} &=& \eta_{\lambda\rho} \slash k -
\frac12 (\gamma_\rho k_\lambda + \gamma_\lambda k_\rho) \\*
\label{qGamma_d2}
k^\mu \Gamma^{(d)}_{\mu\lambda\rho} &=& 0 \,.
\end{eqnarray}
Adding Eqs.\ (\ref{qGamma_a2}) - (\ref{qGamma_d2}) and
using the explicit forms of $V_{\lambda\rho}$ we obtain 
\begin{eqnarray}
k^\mu \Gamma_{\mu\lambda\rho} &=& \eta_{\lambda\rho} (\slash k -
\slash q) \,,
\end{eqnarray}
which vanishes between the spinors. Thus, the amplitude satisfies
electromagnetic gauge invariance as well.

It is interesting to note in this context that, in order to verify the
electromagnetic gauge invariance, there is no need to assume 
that the graviton is on-shell, 
whereas we had to put the photon on-shell to prove
gravitational gauge invariance. The reason for this is the
following. In the proof of a given kind of gauge invariance, only the
particles carrying the corresponding charge need to be on-shell. The
graviton is electromagnetically neutral so that only the electrons
have to be put on-shell in proving electromagnetic gauge
invariance. On the other hand, the electrons as well as the photons
have gravitational couplings; i.e., both are ``gravitationally
charged''.  So, the amplitude is transverse with respect to the
graviton only if the electrons and the photon are all on-shell.

\subsection{Calculation of the cross section}
We will calculate the cross section of the process in the lab frame,
in which the initial electron is at rest, and therefore we set
\begin{eqnarray}
p^\mu = (m,0,0,0) \,.
\end{eqnarray}
In that frame we also write
\begin{eqnarray}
k^\mu & = & (\omega,\vec k) \nonumber\\
q^\mu & = & (\omega^\prime,\vec q) \,.
\end{eqnarray}
The polarization vector $\epsilon^\mu$ of the photon
satisfies
\begin{eqnarray}
\label{photonvectors}
\epsilon^\mu \epsilon_\mu &=& -1 \,,
\end{eqnarray}
in addition to the transversality condition given in 
Eq.\ (\ref{photonshell}).  Since the amplitude is gauge invariant,
we can use the freedom of adding any multiple of
$k^\mu$ to the polarization vector to ensure that its time component
is zero, i.e.,
\begin{eqnarray}
\epsilon^\mu = (0, \vec \epsilon) \,,
\label{eps}
\end{eqnarray}
which defines the {\em radiation gauge}. 
In the present context, this choice of gauge ensures that 
\begin{eqnarray}
p \cdot \epsilon = 0 \,,
\label{transph}
\end{eqnarray}
which will be very useful to simplify the expression for the amplitude
encountered below. 

Similarly, the polarization tensor $\varepsilon^{\lambda\rho}$ for the
graviton is a symmetric tensor which satisfies the conditions
\begin{eqnarray}
\varepsilon^{\lambda\rho} \varepsilon_{\lambda\rho} &=& 1 \,, 
\label{varep.varep}\\*
\varepsilon^{\lambda\rho} (q) q_\lambda = 
\varepsilon^{\lambda\rho} (q) q_\rho &=& 0 \,,
\label{transverse}\\*
\varepsilon^{\lambda\rho}  \eta_{\lambda\rho} &=& 0 \,. 
\label{traceless}
\end{eqnarray}
In addition, we can choose the `radiation gauge' defined by
\begin{eqnarray}
\varepsilon^{0\rho} = \varepsilon^{\lambda 0} &=& 0 \,, 
\label{varepsradgauge}
\end{eqnarray}
so that
\begin{eqnarray}
p_\lambda \varepsilon^{\lambda\rho} = 
p_\rho \varepsilon^{\lambda\rho} 
&=& 0 \,.
\label{pEps}
\end{eqnarray}
With these definitions, we can now write down the simplified form for
the various terms in $\cal M$ as defined in Eq.~(\ref{defGamma}),
which we label with
the same letter that labels the corresponding contribution to
$\Gamma_{\mu\lambda\rho}$.

Since the $\eta_{\lambda\rho}$ terms do not contribute to the physical
amplitude due to Eq.~(\ref{traceless}), we can write
\begin{eqnarray}
{\cal M}_a &=& \frac14 \overline u(p') \left[ (2p'+q)_\lambda
\gamma_\rho + (2p'+q)_\rho \gamma_\lambda \right] {1\over \slash p +
\slash k - m} \gamma_\mu u(p)
\varepsilon^{\ast\lambda\rho} \epsilon^\mu \nonumber\\*
& = & \overline u(p') p^\prime_\lambda \slash \varepsilon^{\ast\lambda}
{\slash p +
\slash k + m \over (p+k)^2-m^2} \slash \epsilon u(p) \,,
\end{eqnarray}
where we have used Eq.~(\ref{transverse}) and the shorthand notation
\begin{eqnarray}
\slash \varepsilon^{\ast\lambda} \equiv \gamma_\rho
\varepsilon^{\ast\lambda\rho} = \gamma_\rho
\varepsilon^{\ast\rho\lambda} \,.
\end{eqnarray}
On the other hand,
\begin{eqnarray}
\label{msubarelation}
(\slash p + m) \slash \epsilon u(p) = [2p\cdot \epsilon +
\slash
\epsilon (-\slash p +m)] u(p) =0 \,,
\end{eqnarray}
where the last equality follows from
(\ref{transph}) and the Dirac equation for the spinor.
Using $(p+k)^2-m^2 = 2m\omega$, it then follows that
\begin{eqnarray}
\label{msuba}
{\cal M}_a &=& {k_\lambda \over 2m\omega} 
\overline u(p')  \slash \varepsilon^{\ast\lambda}
\slash k \slash \epsilon u(p) \,,
\end{eqnarray}
where we have used momentum conservation and Eqs. (\ref{transverse})
and (\ref{pEps}) to replace $p'_\lambda$ by $k_\lambda$.

Regarding ${\cal M}_b$, we find that
\begin{eqnarray}
{\cal M}_b &=& 0 \,,
\end{eqnarray}
since either a $p$ or a $q$ contracts with the graviton polarization
tensor, while
\begin{eqnarray}
\label{Mcsimp}
{\cal M}_c &=& -\, [\overline u(p') \slash \varepsilon^{\ast\lambda} u(p)]
\epsilon_\lambda \,.
\end{eqnarray}
Finally, for ${\cal M}_d$, we note that $C_{\mu\nu\lambda\rho}$,
defined in Eq.~(\ref{C}), can be written as
\begin{eqnarray}
C_{\mu\nu\lambda\rho}(k,k-q) = 2\left( -\eta_{\mu\nu} k_\lambda k_\rho
+ \eta_{\lambda\mu} k_\nu k_\rho - \eta_{\lambda\nu} q_\mu k_\rho +
\eta_{\lambda\mu} \eta_{\rho\nu} k\cdot q \right) + \cdots\,,
\end{eqnarray}
where the dots indicate terms which vanish after contraction with the
polarization factors, either because they are antisymmetric in the
indices $\lambda$ and $\rho$, or because of various on-shell
conditions described above.  Therefore,
\begin{eqnarray}
\label{Mdsimp}
{\cal M}_d &=& {1\over k\cdot q} \overline u(p') \gamma^\nu u(p) 
\left( -\eta_{\mu\nu} k_\lambda k_\rho
+ \eta_{\lambda\mu} k_\nu k_\rho - \eta_{\lambda\nu} q_\mu k_\rho +
\eta_{\lambda\mu} \eta_{\rho\nu} k\cdot q \right)
\varepsilon^{\ast\lambda\rho} \epsilon^\mu \,.
\end{eqnarray}
If we add ${\cal M}_c$ and ${\cal M}_d$
the last term in Eq.\ (\ref{Mdsimp}) cancels ${\cal M}_c$, so that
\begin{eqnarray}
{\cal M}'_d \equiv {\cal M}_c + {\cal M}_d &=& 
{1\over k\cdot q} \overline u(p') \gamma^\mu u(p) 
\left[ -\epsilon_\mu k \cdot \zeta^\ast
+ k_\mu \epsilon \cdot \zeta^\ast - q\cdot
\epsilon \zeta^\ast_\mu  
\right] \,,
\end{eqnarray}
where we have defined
\begin{equation}
\label{veczeta}
\zeta^\mu \equiv \varepsilon^{\mu\nu}k_\nu \,.
\end{equation}

We now square the amplitude and sum over the final electron spin and
average over the initial spin. However, we do not sum or average over
the polarizations of the photon nor of the graviton. This gives, for
the different terms,
\begin{eqnarray}
\frac12 \sum_{\rm spin} {\cal M}^*_a {\cal M}_a &=& 
{1\over m\omega} 
\left[ 2| k\cdot\zeta|^2 -
m\omega' (\zeta\cdot\zeta^\ast) \right] \,, \\*
\frac12 \sum_{\rm spin} {\cal M'}^*_d {\cal M}_d &=& 
{2\over (k\cdot q)^2} \left[
2m^2\omega\omega' |\epsilon \cdot \zeta^\ast|^2 +
m(\omega-\omega') \left\{ |k \cdot\zeta| ^2 - (q\cdot\epsilon)^2 (
\zeta\cdot\zeta^\ast) \right\} \right] \\*
\frac12 \sum_{\rm spin} {\cal M}^*_a {\cal M'}_d &=& 
{1\over m\omega k\cdot q} \left[ m\omega (q\cdot \epsilon)^2 
(\zeta\cdot\zeta^\ast) - 2m^2\omega\omega' |\epsilon \cdot \zeta^\ast|^2  
- m (2\omega - \omega') |k \cdot \zeta|^2 
\right] \\*
\frac12 \sum_{\rm spin} {\cal M'}^*_d {\cal M}_a &=& 
\frac12 \sum_{\rm spin} {\cal M}^*_a {\cal M'}_d \,,
\end{eqnarray}
where we have taken $\epsilon^\mu$ to be real, which corresponds to
a linearly polarized photon.
Adding these terms and using the relation
$k\cdot q = m(\omega - \omega^\prime)$, we obtain
\begin{eqnarray}
\overline{|{\cal M}|^2} \equiv \frac12 \sum_{\rm spin} {\cal M}^* {\cal M} &=& 
{4 \omega'^2\over (\omega - \omega')^2} |\epsilon \cdot \zeta^\ast|^2
- {\omega'\over \omega} (\zeta\cdot\zeta^\ast)\,.
\end{eqnarray}
Using the conditions given in Eqs.\
(\ref{eps}) and (\ref{varepsradgauge}), it can be written as
\begin{eqnarray}
\overline{|{\cal M}|^2} & = &
{4 \omega'^2\over (\omega - \omega')^2} |\vec\epsilon\cdot\vec\zeta^\ast|^2 
+ {\omega'\over \omega} |\vec\zeta|^2 \,,
\end{eqnarray}
where, according to Eq.\ (\ref{veczeta})
\begin{eqnarray}
\label{vecv}
\zeta^i = \varepsilon^{ij}k_j \,.
\end{eqnarray}

In order to obtain the differential cross section for an
unpolarized beam of photons, we must to sum over photon polarizations
and divide by 2. The sum over the polarizations is carried out by using
\begin{eqnarray}
\sum \epsilon_i^\ast(k) \epsilon_j(k) = \delta_{ij} - 
\widehat k_i \widehat k_j \,,
\end{eqnarray}
where $\widehat k$ denotes the unit vector in the direction of $\vec
k$.  In this way, the formula that we obtain for the squared
amplitude, summing over the final electron spin and averaging over the
initial electron and photon polarizations is
\begin{eqnarray}
\overline{|{\cal M}|^2}
& = & \frac{\omega^\prime}{\omega}|\vec \zeta|^2 + 
\frac{2\omega^{\prime\, 2}}{(\omega - \omega^\prime)^2}
\left(|\vec \zeta|^2 - \frac{|\vec k\cdot\vec \zeta|^2}{\omega^2}\right) \,.
\end{eqnarray} 

Further, we can represent the graviton polarization tensor by
\begin{eqnarray}
\varepsilon^{\mu\nu} (q) = \epsilon^\nu(\vec q)\epsilon^\nu(\vec q) \,,
\label{epseps}
\end{eqnarray}
where
\begin{eqnarray}
\epsilon^\mu(\vec q) = (0,\vec\epsilon(\vec q))
\end{eqnarray}
are the spin-1 polarization vectors for a definite helicity ($\pm$).
Then it follows that
\begin{eqnarray}
|\vec k\cdot\vec \zeta|^2 = \omega^4 |\widehat k\cdot\vec\epsilon(\vec q)|^4
\nonumber\\ 
|\vec \zeta|^2 = \omega^2 |\widehat k\cdot\vec\epsilon(\vec q)|^2 \,,
\end{eqnarray}
so that
\begin{eqnarray}
\overline{|{\cal M}|^2} & = &
\omega^\prime \omega |\widehat k\cdot\vec\epsilon(\vec q)|^2 + 
\frac{2\omega^{\prime\, 2}\omega^2}{(\omega - \omega^\prime)^2}
\left(|\widehat k\cdot\vec\epsilon(\vec q)|^2 - 
|\widehat k\cdot\vec\epsilon(\vec q)|^4\right) \,.
\end{eqnarray}

The differential cross section is given by 
\begin{eqnarray}
{d\sigma \over d\Omega} = \frac{\alpha G}{2m^2}\left(\frac{\omega^\prime}{\omega}\right)^2
\overline{|{\cal M}|^2}
\end{eqnarray}
where
\begin{eqnarray}
\omega^\prime = \frac{\omega}{1 + \frac{\omega}{m}(1 - \cos\theta)}\,,
\label{omega'}
\end{eqnarray}
$\theta$ being the angle between $\vec q$ and $\vec k$.

Let us write the components of $\widehat q$ in the form
\begin{eqnarray}
\widehat q = (\sin\theta\cos\phi,\sin\theta\sin\phi,\cos\theta) 
\end{eqnarray}
in a system of axis where $\vec k$ is along the $z$-direction.
Then
\begin{eqnarray}
\vec\epsilon_\pm = \frac{1}{\sqrt{2}}(\vec\epsilon_1 \pm
i\vec\epsilon_2) 
\end{eqnarray}
where
\begin{eqnarray}
\vec\epsilon_1 & \equiv & (-\sin\phi,\cos\phi,0) \nonumber\\
\vec\epsilon_2 & \equiv & 
(\cos\theta\cos\phi,\cos\theta\sin\phi,-\sin\theta) \,.
\end{eqnarray}
{}From these relations we then have
\begin{eqnarray}
|\widehat k\cdot\vec\epsilon(\vec q) |^2 = \frac{1}{2}\sin^2\theta \,,
\end{eqnarray}
and finally
\begin{eqnarray}
\overline{|{\cal M}|^2} = 
\frac12 \left[ 
\omega \omega'  \left(1 - \cos^2\theta\right) + 
\frac{\omega^2 \omega'\,^2}{(\omega - \omega')^2}
\left(1 - \cos^4\theta\right) \right]\,,
\end{eqnarray}
for either helicity of the graviton.  Summing over the graviton
helicities, the total differential cross section is then given by
\begin{eqnarray}
\label{difsigmacompton}
\frac{d\sigma_{tot}}{d(\cos\theta)} = 
\frac{\pi\alpha G}{m^2} 
\omega^{\prime\,^2} \; \left[ 
{\omega' \over \omega}  \left(1 - \cos^2\theta\right) + 
\frac{\omega'\,^2}{(\omega - \omega')^2}
\left(1 - \cos^4\theta\right) \right] \,,
\end{eqnarray}
where $\omega'$ depends on $\cos\theta$ through Eq.\
(\ref{omega'}).  The second term in square brackets in
Eq.\ (\ref{difsigmacompton}) exhibits the singularity
in the forward direction
that is typical of the Coulomb potential,
which is due to its long range character.
In the present case, it has made its way through
the photon exchange diagram of Fig.\ \ref{f:photprod}.

\section{Gravity from pair annihilation}
\setcounter{equation}{0}
\label{s:pairann}
We now consider the related process
\begin{eqnarray}
e^- (p) + e^+ (p') \to \gamma(k) + {\cal G} (q) \,,
\end{eqnarray}
where a graviton and a photon are produced due to the annihilation of
an electron-positron pair.  The amplitude for this process is obtained
from Eq.\ (\ref{defGamma}) by making the substitutions
\begin{eqnarray}
\label{4momcrossing}
p^\prime & \rightarrow&  -p^\prime\,,\qquad  k \rightarrow -k \\
\label{wfcrossing}
u(p') & \rightarrow & v(p')\,, \qquad
\epsilon^{\mu}\rightarrow \epsilon^{\ast\mu} \,.
\end{eqnarray}
If we write the amplitude in the form
\begin{eqnarray}
iM &  = & (-i\kappa)(ie)
i\left[\overline v(p')\Gamma_{\mu\lambda\rho}u(p) \right]
\varepsilon^{\ast\lambda\rho}(q) \epsilon^{\ast\mu}(k) \nonumber\\
& \equiv & ie\kappa{\cal M} \,,
\label{defGamma'}
\end{eqnarray}
the contributions to $\Gamma_{\mu\lambda\rho}$ from the various
diagrams are 
\begin{eqnarray}
\Gamma^{(a)}_{\mu\lambda\rho} &=& V_{\lambda\rho}(q-p',-p')
S_F(q-p') \gamma_\mu \\ 
\Gamma^{(b)}_{\mu\lambda\rho} &=& \gamma_\mu S_F(p-q) 
V_{\lambda\rho} (p,p-q) \\ 
\Gamma^{(c)}_{\mu\lambda\rho} &=& \gamma^\alpha
a_{\alpha\mu\lambda\rho} \\ 
\Gamma^{(d)}_{\mu\lambda\rho} &=&
-\, \gamma^\nu \Delta(-k-q)C_{\mu\nu\lambda\rho} (-k,-k-q)  \,,
\end{eqnarray}
with the notation for vertices and propagators used in the previous
sections. The explicit verification of gauge invariance, 
both gravitational and electromagnetic, is similar to the exercise of
Sec.~\ref{s:photprod}. 

The on-shell relations
\begin{eqnarray}
k^2 = q^2 & = & 0 \nonumber\\
p^2 = p^{\prime\,2} & = & m^2 \,,
\end{eqnarray}
together with Eq.\ (\ref{traceless}),
allow us to write the corresponding contributions 
to ${\cal M}$ as
\begin{eqnarray}
\label{M'a}
{\cal M}_a &=& {p'_\lambda \over 2p \cdot k} 
\left[ \overline v(p') \gamma_\rho (\slash q - \slash p'+m) \gamma_\mu
u(p) \right] \varepsilon^{\ast\lambda\rho}
\epsilon^{\ast\mu} \,, \\
\label{M'b}
{\cal M}_b &=& -\, {p_\lambda \over 2p^\prime \cdot k} 
\left[ \overline v(p') \gamma_\mu (\slash p - \slash q +m) \gamma_\rho
u(p) \right] \varepsilon^{\ast\lambda\rho}
\epsilon^{\ast\mu} \,, \\
\label{M''d}
{\cal M}^{\prime}_d \equiv {\cal M}_c + {\cal M}_d &=& 
{k_\lambda \over k \cdot q} 
\left[ \overline v(p') \gamma^\nu u(p) \right] 
\left( \eta_{\mu\nu} k_\rho - \eta_{\mu\rho} k_\nu - \eta_{\nu\rho}
q_\mu \right) 
\varepsilon^{\ast\lambda\rho} \epsilon^{\ast\mu} \,.
\end{eqnarray}

We will calculate the cross section in the center of mass frame
of the initial particles.  The following considerations then
simplify the expression for the amplitude.
Since
\begin{eqnarray}
p + p' = (\sqrt{s},\vec 0) 
\end{eqnarray}
in that frame, then in the radiation gauge for the graviton
it follows that
\begin{eqnarray}
\label{4kdotvareps}
k_\lambda\varepsilon^{\lambda\rho} = 0
\end{eqnarray}
and
\begin{eqnarray}
\label{4pdotvareps}
p'_\lambda\varepsilon^{\lambda\rho} =
-p_\lambda\varepsilon^{\lambda\rho} \,,
\end{eqnarray}
with analogous identities for the contractions with respect
to the index $\rho$. Eq.\ (\ref{4kdotvareps}) tells us
that ${\cal M}^{\prime}_d = 0$.  The expressions for
${\cal M}_{a,b}$ can be reduced further with the help
of the relations
\begin{eqnarray}
\slash{p}u & = & mu \nonumber\\
\overline v\slash{p}\,' & = & -m\overline v \,.
\end{eqnarray}
In this way we finally obtain
\begin{eqnarray}
\label{4ampfinal}
{\cal M} = {\cal M}_1 + {\cal M}_2 \,,
\end{eqnarray}
with
\begin{eqnarray}
\label{4m12}
{\cal M}_1 & = & \left[\frac{p'\cdot\epsilon^\ast}{p'\cdot k}
- \frac{p\cdot\epsilon^\ast}{p\cdot k}\right]
(\overline v\slash\xi^\ast u) \nonumber\\
{\cal M}_2 & = & \frac{1}{2p\cdot k} \;
(\overline v\slash\xi^\ast\slash{k}\slash{\epsilon}^\ast u) -
\frac{1}{2p'\cdot k} \;
(\overline v\slash{\epsilon}^\ast\slash{k}\slash{\xi}^\ast u) \,.
\end{eqnarray}
where we have defined the vector
\begin{eqnarray}
\xi^\mu \equiv \varepsilon^{\mu\nu}p_\nu \,.
\end{eqnarray}
Notice that while we have chosen a specific gauge for the graviton
polarization tensor in order to arrive at Eq.\ (\ref{4m12}), the gauge
for the photon polarization vector has been left unspecified.  In
particular, the expressions given in Eq.\ (\ref{4m12}) become zero if
the replacement $\epsilon_\mu\rightarrow k_\mu$ is made.  Therefore,
in the square of the amplitudes, the sum over the polarizations of the
photon can be carried out by means of the formula
\begin{eqnarray}
\label{photpolsum}
\sum \epsilon^{\mu}\epsilon^{\ast\nu} = - \eta^{\mu\nu} \,. 
\end{eqnarray}

Thus, taking the square of the amplitude in Eq.\ (\ref{4ampfinal}),
and averaging over the initial electron and positron spins and
summing over the photon polarizations, we obtain
\begin{eqnarray}
\label{4ampsq}
\overline{|{\cal M}_1|}^{\, 2} & = &
\left(\frac{m^2}{(p' \cdot k)^2} +
\frac{m^2}{(p\cdot k)^2} -
\frac{2p\cdot p'} {(p\cdot k)(p'\cdot k)}
\right)
\left[
2(p\cdot\xi)(p\cdot\xi^\ast) + p\cdot p' (\xi\cdot\xi^\ast) + 
m^2(\xi\cdot\xi^\ast)
\right]
\nonumber\\
\overline{|{\cal M}_2|}^{\, 2} & = &
-\left(
\frac{p\cdot k}{p'\cdot k} +
\frac{p' \cdot k}{p\cdot k}
\right)(\xi\cdot\xi^\ast)
\nonumber\\
2\overline{{\cal M}_1{\cal M}_2^\ast} & = &
-\left(
\frac{p'\cdot k}{(p\cdot k)^2} +
\frac{p\cdot k}{(p'\cdot k)^2}\right)m^2|\xi|^2 \nonumber\\
& & \mbox{} +
\left(
\frac{1}{p\cdot k} + \frac{1}{p'\cdot k}
\right)
\left[
-m^2(\xi\cdot\xi^\ast) + 2p\cdot p'(\xi\cdot\xi^\ast) + 
4(p\cdot\xi)(p\cdot\xi^\ast)
\right]\,.
\end{eqnarray}
Adding these, and using the relation
\begin{eqnarray}
p\cdot p' + m^2 = (p + p')\cdot k \,,
\end{eqnarray}
the total amplitude squared is then given by
\begin{eqnarray}
\label{4totampsq}
\overline{|{\cal M}|^2} = 
2m^2\left(
\frac{1}{p\cdot k} + \frac{1}{p'\cdot k}
\right)^2 (p\cdot\xi)(p\cdot\xi^\ast) -
\left(
\frac{1}{p\cdot k} + \frac{1}{p'\cdot k}
\right)(\xi\cdot\xi^\ast) \,.
\end{eqnarray}

To proceed further, we make a construction of the 
graviton polarization tensor in a way analogous
to that given in Eq.\ (\ref{epseps}). 
Thus, introducing the angle $\theta$ by writing
\begin{eqnarray}
\cos\theta\equiv \widehat q\cdot\widehat p \,,
\end{eqnarray}
we have
\begin{eqnarray}
(p\cdot\xi)(p\cdot\xi^\ast) & = & \frac{1}{4}|\vec p|^4\sin^4\theta 
\nonumber\\
\xi\cdot\xi^\ast & = & -\frac{1}{2}|\vec p|^2\sin^2\theta \,,
\end{eqnarray}
where $|\vec p|$ is the magnitude of the electron momentum, given by
\begin{eqnarray}
|\vec p| = \frac12\beta\sqrt{s} \,,
\end{eqnarray}
with 
\begin{eqnarray}
\beta = \sqrt{1 - \frac{4m^2}{s}} 
\end{eqnarray}
being the electron velocity. In addition, we use the kinematic formulas
\begin{eqnarray}
p\cdot k & = & \frac{1}{4}s(1 + \beta\cos\theta) \nonumber\\
p'\cdot k & = & \frac{1}{4}s(1 - \beta\cos\theta) \,.
\end{eqnarray}
Substituting these in Eq.\ (\ref{4totampsq})
we finally arrive at
\begin{eqnarray}
\label{4ampsqexp}
\overline{|{\cal M}|^2} & = & \frac12 m^2\beta^4\sin^4\theta
\left[\frac{1}{1 + \beta\cos\theta} + \frac{1}{1 - \beta\cos\theta}
\right]^2 \nonumber\\[12pt]
& & \mbox{}  + 
\frac18 s\beta^2\sin^2\theta 
\left[
\frac{1 - \beta\cos\theta}{1 + \beta\cos\theta} +
\frac{1 + \beta\cos\theta}{1 - \beta\cos\theta}
\right] \,,
\end{eqnarray}
which holds for either helicity of the graviton.  
If we sum over the graviton helicities, we must multiply
Eq.\ (\ref{4ampsqexp}) by a factor of 2.
The total differential cross section, summed over the
graviton helicities, is then given by
\begin{eqnarray}
\label{annsigmatot}
\frac{d\sigma_{\rm tot}}{d(\cos\theta)} & = &
\frac{\pi G\alpha\beta}{s}\left\{
m^2\beta^2\sin^4\theta
\left[\frac{1}{1 + \beta\cos\theta} + \frac{1}{1 - \beta\cos\theta}
\right]^2\right. \nonumber\\[12pt]
& & \mbox{}  + 
\frac{s}{4}\sin^2\theta 
\left.
\left[
\frac{1 - \beta\cos\theta}{1 + \beta\cos\theta} +
\frac{1 + \beta\cos\theta}{1 - \beta\cos\theta}
\right]
\right\} \,.
\end{eqnarray}

It is curious to observe that, at energies of the order of
$\sqrt{s}\simeq 10^{17}$\,GeV, such as those that may have been
available in the Early Universe, $\sigma_{\rm tot}$ becomes of the
order of a few percent of a typical cross section $\sigma_0\simeq
(\alpha^2/s)$ for producing photons and the weak gauge bosons in
analogus reactions.
Of course, at such energies the non-linear terms in the
gravitational interactions might play a significant role and
change the cross section appreciably.

\section{Conclusions}
\label{s:conc}
Unless our experience and current understanding has totally misled us,
it is likely that, at some level, the graviton participates actively
in physical processes as a real particle and not just as a mediator of
the gravitational interactions.

In this work we have considered in some detail the calculation of the
cross section for two such processes, in which a graviton emerges in
the final state, and which involve both the gravitational and
electromagnetic interactions.  Special attention has been put in the
derivation of the appropriate couplings that should be used in the
Lagrangian, and in showing that with the proper choice of them the
amplitudes for the processes are gauge invariant.

Apart from the intrinsic interest and from any direct application that
these calculations might have, they provide a useful setting to study
some of the subtle technical issues that are involved in this type of
calculation, disentangled from the additional complications that
appear in the study of processes that occur in a background medium.
As already remarked in the Introduction and commented upon in Section\
\ref{s:int}, the details of the present work are relevant for the kind
of calculation carried out in Ref.\ \cite{np:gravnu}, which have to do
with the matter-induced gravitational interactions of neutrinos and
other particles in a background medium.  These calculations, which
extend the work of Ref.\ \cite{np:gravnu}, are currently in progress.


\end{document}